\input{aipcheck}

\documentclass[
  ,draft            
  ]
  {aipproc}

\layoutstyle{6x9}
\RequirePackage{xspace}
\RequirePackage{amsmath}

\RequirePackage{amsmath}
\RequirePackage{amsfonts}
\RequirePackage{amssymb}

\def\ifundefined#1{\expandafter\ifx\csname#1\endcsname\relax}

\def\la{\mathrel{\hbox{\rlap{\hbox{\lower4pt\hbox{$\sim$}}}\hbox{$<$}}}}
\def\ga{\mathrel{\hbox{\rlap{\hbox{\lower4pt\hbox{$\sim$}}}\hbox{$>$}}}}

\newcommand{\be}{\begin{eqnarray}}
\newcommand{\ee}{\end{eqnarray}}

\ifundefined{ensuremath}\def\ensuremath#1{\relax\ifmmode{#1}}
\else${#1}$\fi\else\relax\fi
\ifundefined{nuc}\def\nuc#1#2{\relax\ifmmode{}^{#1}{\protect\textrm{#2}}
\else${}^{#1}$#2\fi}\else\relax\fi

\newcommand{\etal}{et al.\xspace}

\newcommand{\kmps}{\ensuremath{\mathrm{km}~\mathrm{s}^{-1}}\xspace}

\def\Tmod{\ensuremath{T_{\mathrm{model}}}\xspace}
\def\Teff{\ensuremath{T_{\mathrm{model}}}\xspace}
\def\teff{\ensuremath{T_{\mathrm{model}}}\xspace}
\def\tstd{\ensuremath{\tau_{\mathrm{std}}}\xspace}

\newcommand{\vno}{\ensuremath{v_0}\xspace}

\newcommand{\phx}{\texttt{PHOENIX}\xspace}

\newcommand{\hbeta}{H$\beta$\xspace}

\begin{document}

\title{Spectral Modeling of Type II Supernovae}


\author{E. Baron}{
  address={Homer L.~Dodge Dept.~of Physics and Astronomy, University of
Oklahoma, 440 West Brooks, Rm.~100, Norman, OK 73019, USA},
altaddress={Computational Research Division, Lawrence Berkeley
  National Laboratory, MS 50F-1650, 1 Cyclotron Rd, Berkeley, CA
  94720-8139 USA}
}

\begin{abstract}
Using models of the SN~IIP 2005cs, we show that detailed spectral
analysis can be used to determine reddening and abundances.
\end{abstract}

\maketitle


\section{Introduction}

One of the primary goals of studying supernova spectra is to
understand the details of stellar evolution at the end of a star's
life. Massive stars will produce iron white dwarf cores, which grow
above their Chandrasekhar mass and core-collapse to produce supernovae
and sometimes gamma-ray bursts. Detailed analysis of the spectrum of
the supernova over a wide range of wavelength and time provides a
window into the makeup of the star prior to explosion as well as
details of the explosion process itself and the nature of the
circumstellar medium.

\section{Quantitative Spectroscopy}

\begin{figure}
\centering
\includegraphics[height=7cm,width=7cm,angle=0,clip]{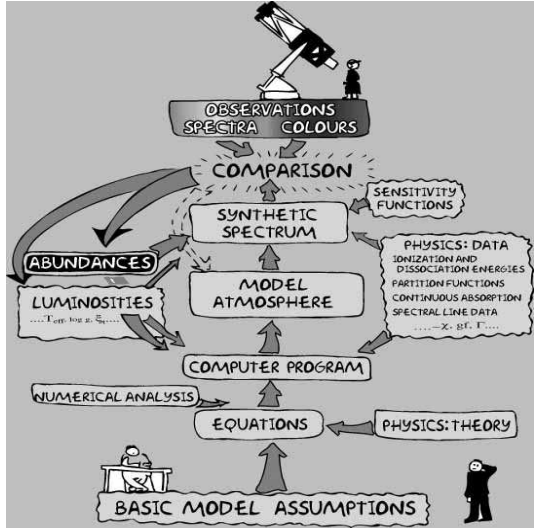}
\caption{\label{fig:dianna}A cartoon that indicates the raison d'\^etre
  for quantitative spectroscopy and the methodology}
\end{figure}

Figure~\ref{fig:dianna} shows the methodology of quantitative
spectroscopy in cartoon form. The basic goal is to produce detailed
NLTE synthetic spectra, compare them to observations and then use
those results to compare to theoretical predictions about the endpoint
of stellar evolution models and explosions. 
We describe calculations performed using the multi-purpose stellar
atmospheres program \phx~{version \tt 14}
\citep{hbjcam99,bhpar298,hbapara97,phhnovetal97,phhnovfe296}.
\phx\ solves the radiative transfer equation along characteristic rays
in spherical symmetry including all special relativistic effects.  The
non-LTE (NLTE) rate equations for many ionization states are solved
including the effects of ionization due to non-thermal electrons from
the $\gamma$-rays produced by the  radiative decay of $^{56}$Ni, which
is produced in the
supernova explosion.  For most of the calculations presented in this
paper the atoms and ions calculated in NLTE are: H~I,
He~I--II, C~I-III, N~I-III, O~I-III, Ne~I, Na~I-II, Mg~I-III, Si~I--III,
S~I--III, Ca~II, Ti~II, Fe~I--III, Ni~I-III, and Co~II. These are all the
elements whose features make important contributions to the observed
spectral features in SNe~II.

Each model atom includes primary NLTE transitions, which are used to
calculate the level populations and opacity, and weaker secondary LTE
transitions which are are included in the opacity and implicitly
affect the rate equations via their effect on the solution to the
transport equation \citep{hbjcam99}.  In addition to the NLTE
transitions, all other LTE line opacities for atomic species not
treated in NLTE are treated with the equivalent two-level atom source
function, using a thermalization parameter, $\alpha =0.05$.  The
atmospheres are iterated to energy balance in the co-moving frame;
while we neglect the explicit effects of time dependence in the
radiation transport equation, we do implicitly include these effects,
via explicitly including the rate of gamma-ray deposition in the generalized
equation of radiative equilibrium and in the rate equations for the
NLTE populations.

The models are parameterized by the time since explosion and the
velocity where the continuum optical depth in extinction at
5000~\AA\ ($\tstd$) is unity, which along with the density profile
determines the radii. This follows since the explosion becomes
homologous ($v \propto r$) quickly after the shock wave traverses the
entire star. The density profile is taken to be a power-law in radius:
\[ \rho \propto r^{-n} \]
where $n$ typically is in the range $6-12$. Since we are only modeling
the outer atmosphere of the 
supernova, this simple parameterization agrees well with detailed
simulations of the light curve \citep{blinn87a00} for the relatively
small regions of the ejecta that our models probe.

Further fitting parameters are the model temperature $\Tmod$, which is
a convenient way of parameterizing the total luminosity in the
observer's frame. We treat the $\gamma$-ray deposition in a simple
parameterized way, which allows us to include the effects of nickel
mixing which is seen in nearly all SNe~II. Here we present preliminary
results of modeling the nearby, well-observed, SN IIP 2005cs. More
detailed results are presented in Ref.~\citep{b05cs06}.

\section{Reddening}

\begin{figure}[ht]
\centering
\includegraphics[height=0.35\textheight,angle=90]{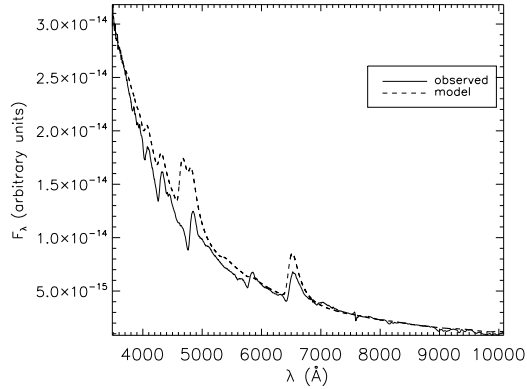}
\caption{\label{fig:sol_t18_ebmv11}Day 5: A synthetic spectrum using
  E(B-V) = 0.11, \Teff = 18,000~K is compared to the observation.}
\end{figure}

\begin{figure}[ht]
\centering
\includegraphics[height=0.35\textheight,angle=90]{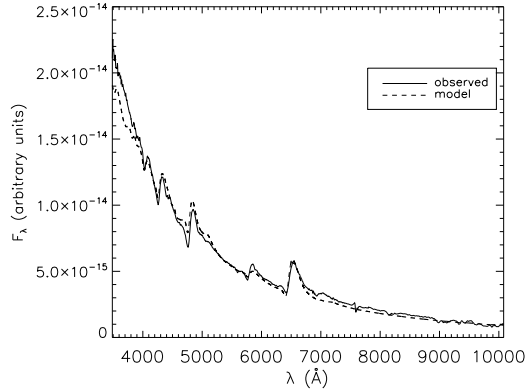}
\caption{\label{fig:sol_t12_ebmv035}Day 5: A synthetic spectrum using E(B-V)
  = 0.035, \Teff = 12,000~K is compared to the observation.} 
\end{figure}

Determining the extinction to SNe~II is difficult, since they are such
a heterogeneous class, it is difficult to find an intrinsic feature
in the spectrum or light curve that can be used to find the parent
galaxy extinction. \citet{sn93w103,bsn99em00} found that the Ca II H+K
lines can be used as a temperature indicator in modeling very early
observed spectra. For SN~2005cs the reddening has been estimated in a
variety of ways. \citet{maund05cs05} used the relationship between the
equivalent width of the Na~I~D interstellar absorption line to obtain a
color excess of $E(B-V) = 0.16$, as well as the color magnitude
diagram of red supergiants within 2 arcsec of SN~2005cs to obtain
$E(B-V) = 0.12$, and their final adopted $E(B-V) =
0.14$. \citet{li05cs06} noted the large scatter in the relationships
for equivalent width of the Na~I~D line, obtaining a range of $E(B-V)
= 0.05-0.20$. Assuming that the color evolution of SN~2005cs is
similar to that of SN~1999em, they found $E(B-V) = 0.12$. Also using
the Na~I~D  line \citet{pasto05cs06} found $E(B-V) = 0.06$, but noting
the uncertainty and comparing with the work of other authors they
adopted $E(B-V) = 0.11$. We began our work by adopting the reddening
estimate of \citet{pasto05cs06}, since our spectra were obtained from
these authors. 
Figure~\ref{fig:sol_t18_ebmv11} shows our best fit using solar
abundances \citep{GS98} where the observed spectrum has been
dereddened using the 
reddening law of \citet{card89} and
$R_V=3.1$. It is evident that the 
region around H$\beta$\ is very poorly fit, there is a strong feature
just to the blue of \hbeta\ and \hbeta\ itself is far too weak. We
attempted to alter the model in a number of ways, changing the density
profile, velocity at the photosphere, and gamma-ray deposition in
order to strengthen \hbeta, however we were unable to find any set of
parameters that would provide a good fit to \hbeta\ (and the rest of
the observed spectrum) with this choice of
reddening. This model has \teff=18000~K, \vno = 6000~\kmps, and $n =
8$. Fig.~\ref{fig:sol_t12_ebmv035}
shows that if \teff\ is reduced to 12000~K and the color excess is
reduced to the galactic foreground value of $E(B-V) = 0.035$
\citep{schlegelred98} the fit is significantly improved. Our value of
$E(B-V) = 0.035$ is in agreement within the errors of the lower values
found by \citet{li05cs06} and \citet{pasto05cs06}. Thus $E(B-V)=
0.035-0.05$, but we will adopt the value of 0.035 for the rest of this
work. This lower value of the extinction will somewhat lower
the inferred mass of the progenitor found by \citet{maund05cs05}
and \citet{li05cs06}, but other uncertainties such as distance and
progenitor metallicity also play important roles in the uncertainty of
the progenitor mass. Clearly the bluest part of the continuum is
better fit with the \teff=18000~K models than with the \teff=12000~K
models. We did not attempt to fine tune our results to perfectly fit
the bluest part of the continuum since the flux calibration at the
spectral edges is difficult and it represents our uncertainty in
\teff\ 
and $E(B-V)$. None of the above models have He~I $\lambda 5876$ strong
enough. It is well known that Rayleigh-Taylor instabilities lead to
mixing between the hydrogen and helium shells, thus our helium
abundance is almost certainly too low, but we will not explore helium
mixing further.

\section{Abundances}

Typically, the approach to obtaining abundances in differentially
expanding flows has been through line identifications. This method has
been extremely successful using the SYNOW code 
\citep[see][and references therein]{branchcomp105,branchcomp206} as
well as the work of Mazzali and collaborators \citep[for
example][]{stehle02bo05}.  
Nevertheless, line identifications do not provide direct information
on abundances, which are what are input into stellar evolution
calculations and output from hydrodynamical calculations of supernova
explosions and nucleosynthesis. Line identifications are subject to
error in that there may be another candidate line that is not
considered in the analysis or there may be two equally valid possible
identifications, the classic being He~I $\lambda 5876$ and Na~I D,
which have very similar rest wavelengths and are both expected in
supernovae (and have both been identified in supernovae).

Here we focus on nitrogen. Massive stars which are the
progenitors of SNe~II are expected to undergo CNO processing at the
base of the hydrogen envelope followed by mixing due to dredge up and
meridional circulation. This would lead to enhanced nitrogen 
and depleted carbon and oxygen. For our CNO processed models we
take the abundances used by \citet{dessart05a}. However, significant
mixing of the hydrogen and helium envelope is expected to occur during
the explosion due to Rayleigh-Taylor instabilities and mass loss will
occur during the pre-supernova evolution. 

\subsection{N~II}

\begin{figure}[ht]
\centering
\includegraphics[height=0.35\textheight,angle=90]{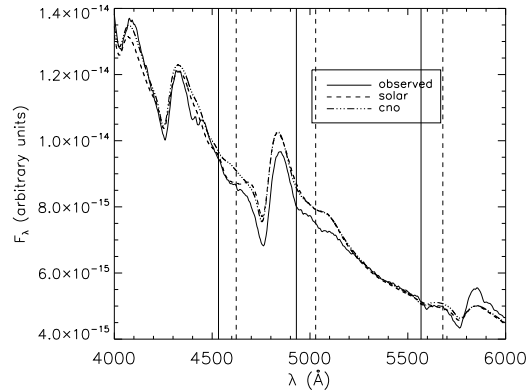}
\caption{\label{fig:cno_v_sol_blowup}A synthetic spectrum using E(B-V)
  = 0.035, \Teff = 12,000~K and enhanced CNO abundances is compared
  the solar model and to
  the observation. A blowup around the region
where optical N~II lines are prominent. The vertical dashed lines show
the rest wavelength for the three lines, the vertical solid lines are
blueshifted by 6000~\kmps.}
\end{figure}

N~II lines were first identified in SN~II  in SN
1990E \citep{bs90E93}. 
Using SYNOW, \citet{bsn99em00} found evidence for N~II in SN~1999em,
however more detailed modeling with \phx\ indicated that the lines were
in fact due to high velocity Balmer and He~I lines.   Prominent N~II
lines in the optical are N~II $\lambda 4623$, $\lambda 5029$, and
$\lambda 5679$. \citet{dessart05a} found strong evidence for N~II in
SN~1999em. Using SYNOW Elmhamdi \etal\ (in preparation) found
evidence for N~II in SN~2005cs, as did
\citet{pasto05cs06} using the code developed by Mazzali and
collaborators. Figure~\ref{fig:cno_v_sol_blowup}
compares solar abundances to a model with enhanced CNO abundances
\citep{dessart05a,prantzos86}. Figure~\ref{fig:cno_v_sol_blowup} shows that
the CNO enhanced abundances model does somewhat better fitting the
emission peak of \hbeta\ (see \S~\ref{sec:line_ids}), however the feature to
the blue of \hbeta, 
clearly well-fit in the solar abundance model is completely absent in
the model with enhanced CNO
abundances. Figure~\ref{fig:cno_v_sol_blowup} shows the region where
the optical N~II lines are prominent and in particular, the N~II
$\lambda 5679$ is not quite in the same place as the observed feature
and the two bluer lines have almost no effect. 

\subsection{Line IDs\label{sec:line_ids}}

\begin{figure}[ht]
\centering
\includegraphics[height=0.35\textheight,angle=90]{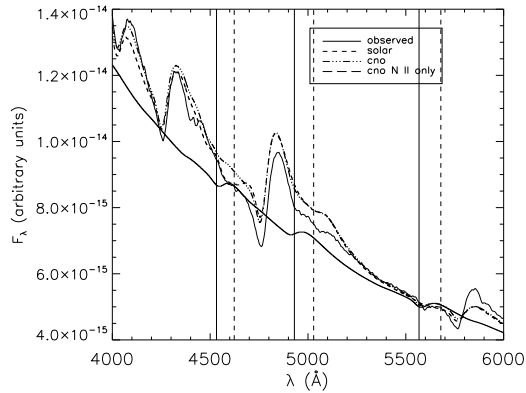}
\caption{\label{fig:nii_single_element}CNO enhanced N~II Single Element}
\end{figure}

In detailed line-blanketed models such as the ones presented here line
identifications are difficult since nearly every feature in the model
spectrum is a blend of many individual weak and strong
lines. Nevertheless, it is useful to attempt to understand just what
species are contributing to the variations in the spectra. In order to
do this we produce ``single element spectra'' where we calculate the
synthetic spectrum (holding the temperature and density structure
fixed) but turning off all line opacity except for that of a given
species. Figure~\ref{fig:nii_single_element} shows the single element
spectrum for N~II for our \teff\ = 12000~K models 
with CNO enhanced abundances. Clearly the N~II lines are
present in CNO enhanced models, but
their effect on the total spectrum is unclear. 

\begin{figure}[ht]
\centering
\includegraphics[height=0.35\textheight,angle=90]{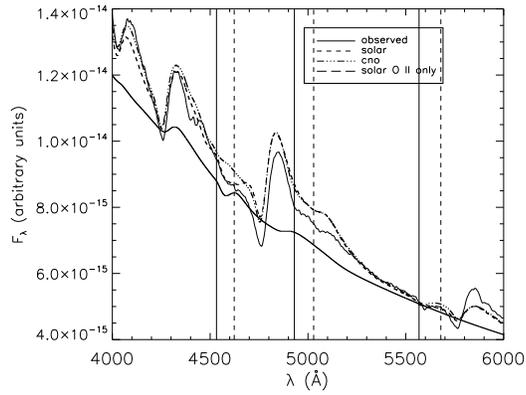}
\caption{\label{fig:oii_single_element}Day 5: Solar O~II Single Element}
\end{figure}

In an attempt to identify the better fit of the feature just to the
blue of \hbeta\ we examined the single element spectrum of 
O~II. Figure~\ref{fig:oii_single_element} clearly shows that O~II
lines play an important role in producing the observed feature just
blueward of \hbeta. Most likely it is the lines O~II $\lambda 4651.5$
and $\lambda 4698$ which are producing the observed feature. On the
other hand it is also clear that O~II $\lambda 4915$ and $\lambda
4943$ are producing the deleterious feature just to the red of
\hbeta. 

Thus, it is clear that the strong depletion of oxygen expected from
CNO processing is not evident, we can not rule out that there is some
enhanced nitrogen in the observed spectra, but the N~II lines don't
seem to form in the right place. However since we are studying simple,
parameterized, homogeneous models this could be an artifact of our
parameterization. Nevertheless, the absorption trough of the feature
that we would like to attribute to N~II $\lambda 5679$ is too fast in
our models, whereas one would expect the N~II to be more enhanced on
the outermost part of the envelope and thus to form at even higher
velocity due to homologous expansion. Clumping could of course change
this simple one-dimensional picture.

\begin{figure}[ht]
\centering
\includegraphics[height=0.35\textheight,angle=90]{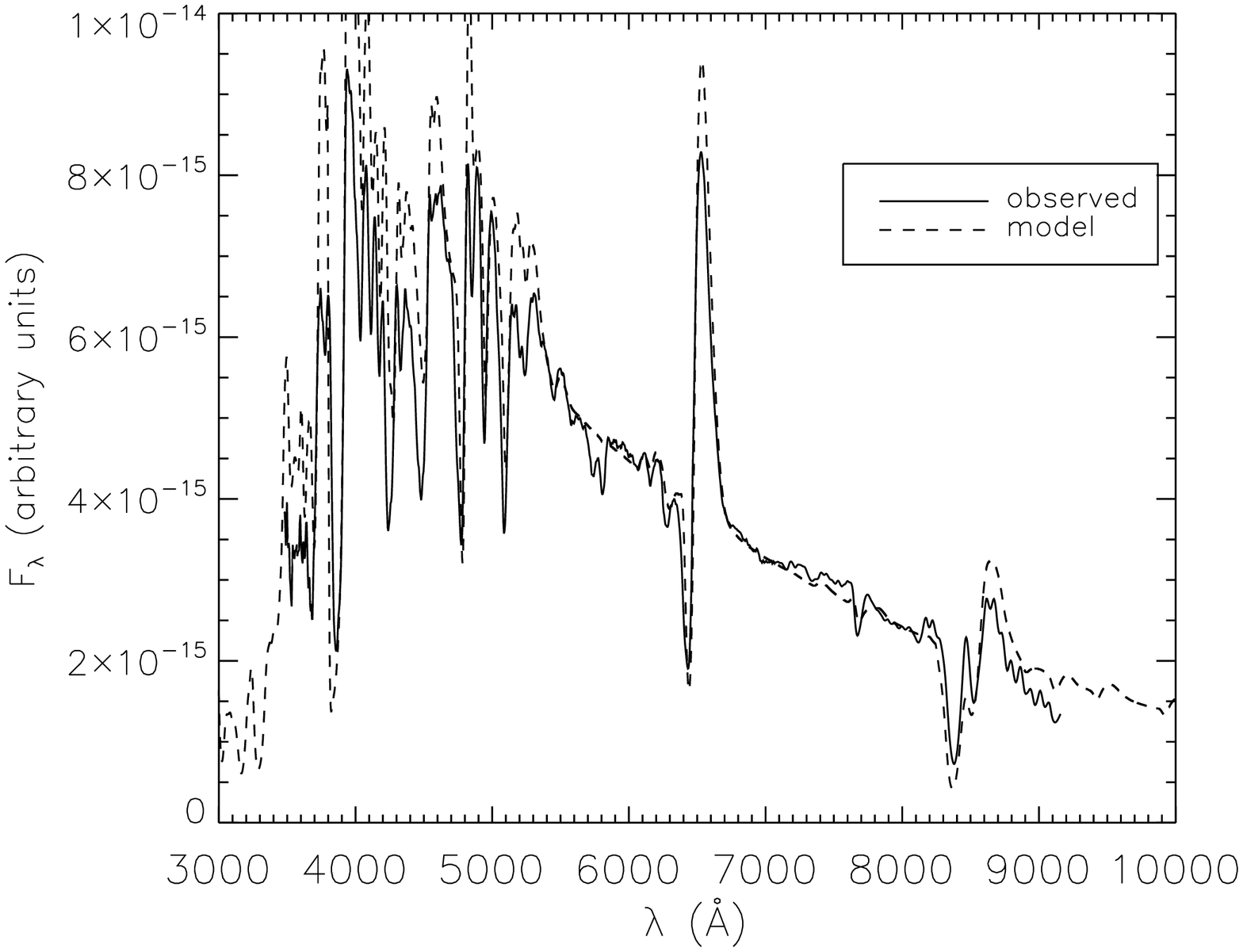}
\caption{\label{fig:day16}Day 17: A preliminary synthetic spectrum is
  compared to observation.}
\end{figure}

\begin{figure}[ht]
\centering
\includegraphics[height=0.35\textheight,angle=90]{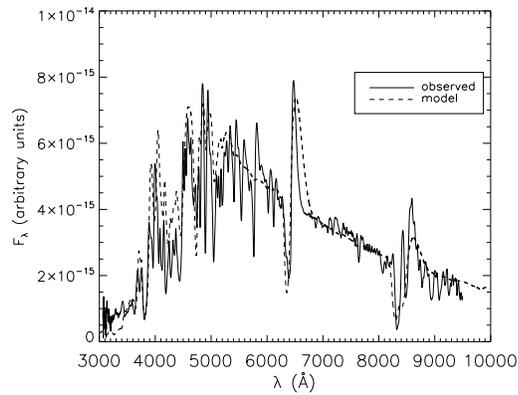}
\caption{\label{fig:day32}Day 34: A preliminary synthetic spectrum is
  compared to observation.}
\end{figure}

Figure~\ref{fig:day16} shows a preliminary synthetic spectrum compared
to the observation 17 days after explosion and Fig.~\ref{fig:day32}
shows the same for 34 days after explosion. Dessart \& Hillier (this
volume) find that time dependence in the rate equations is important
for reproducing the Balmer lines. While our Balmer lines aren't
perfect they are quite reasonable and all relevant physical processes
need to be included in the calculations.

\clearpage


\begin{theacknowledgments}
  We think Andrea Pastorello for providing us with unpublished spectra
  and Abouazza Elmhamdi and Andrea Pastorello for helpful discussions.
  This work was supported in part  by NASA grants NAG5-3505 and
  NNG04GD368, and NSF grant AST-0307323.  
  This research used resources of the National Energy Research
  Scientific Computing Center (NERSC), which is supported by the
  Office of Science of the U.S.  Department of Energy under Contract
  No. DE-AC03-76SF00098; and the H\"ochstleistungs Rechenzentrum Nord
  (HLRN).  We thank all these institutions for a generous allocation
  of computer time.
\end{theacknowledgments}


\end{document}


\endinput